\documentclass[prl,aps,twocolumn, superscriptaddress]{revtex4-2}
\usepackage{natbib}
\usepackage{amsfonts}
\usepackage{amssymb}
\usepackage{amsmath}
\usepackage{graphicx}
\usepackage{verbatim}
\usepackage{hyperref,physics,bm}
\usepackage{lipsum, wrapfig}
\usepackage{enumitem}
\usepackage{hyperref}
\usepackage[dvipsnames]{xcolor}
\usepackage{soul}
\usepackage[normalem]{ulem} 

\hypersetup{pdflang={English},colorlinks=true,linkcolor=Cerulean,citecolor=RoyalBlue,urlcolor=RoyalBlue,breaklinks=true}

\renewcommand{\a}{\hat{a}}
\newcommand{\adag}{\hat{a}^\dag}
\newcommand{\x}{\hat{x}}
\newcommand{\p}{\hat{p}}
\newcommand{\D}{\hat{D}}
\newcommand{\R}{\hat{R}}
\newcommand{\U}{\hat{U}}

\newcommand{\Ddag}{\hat{D}^\dagger}
\renewcommand{\S}{\hat{S}}
\newcommand{\sx}{\hat{\sigma}_x}

\newcommand{\sz}{\hat{\sigma}_z}

\begin{document}
\title{Quantum Sensing Using Geometrical Phase in Qubit-Oscillator Systems}
\author{Nishchay  Suri}\thanks{nsuri@lbl.gov}
\affiliation{Applied Mathematics and Computational Research Division, Lawrence Berkeley National Laboratory, Berkeley, California 94720, USA}
\author{Zhihui Wang}
\affiliation{Quantum Artificial Intelligence Laboratory, NASA Ames Research Center, Moffett Field, CA 94035, USA}
\affiliation{USRA Research Institute for Advanced Computer Science, Mountain View, California 94043, USA}
\author{Tanay Roy}
\affiliation{Superconducting Quantum Materials and Systems (SQMS) Center,
Fermi National Accelerator Laboratory (FNAL), Batavia, IL 60510, USA}
\author{Davide Venturelli}
\affiliation{Quantum Artificial Intelligence Laboratory, NASA Ames Research Center, Moffett Field, CA 94035, USA}
\affiliation{USRA Research Institute for Advanced Computer Science, Mountain View, California 94043, USA}
\author{Wibe Albert de Jong}
\affiliation{Applied Mathematics and Computational Research Division, Lawrence Berkeley National Laboratory, Berkeley, California 94720, USA}
\begin{abstract}
We present a quantum sensing protocol for coupled qubit–oscillator systems that surpasses the standard quantum limit (SQL) by exploiting a geometrical phase. The signal is encoded in the geometrical phase that is proportional to the area enclosed in oscillator phase space. This area is amplified through squeezing, enabling sensitivities beyond the SQL. Our method is independent of oscillator’s initial state, amenable to sensing with high-temperature or logical error-corrected states. The protocol shows robustness to qubit Markovian noise and preserves its state-independence, underscoring its practicality for next-generation quantum metrology. We demonstrate application to force sensing beyond the SQL in longitudinally coupled systems, and to high-precision measurements of couplings and pulse calibration surpassing SQL in dispersively coupled circuit quantum electrodynamics (cQED) architectures.
\end{abstract}

\maketitle

\textit{Introduction---} 
Quantum technologies have seen remarkable progress in recent years, especially in the field of quantum computing and information processing. Qubit–oscillator systems are ubiquitous in this landscape, forming the basis of architectures ranging from superconducting circuits and trapped ions to optomechanical, spin-mechanical, and semiconductor devices~\cite{blais2004cavity, leibfried2003experimental, aspelmeyer2014cavity, kolkowitz2012coherent, lodahl2015interfacing}.
These systems combine the exceptional force sensitivity of the oscillator with the precise control and high-fidelity readout of the qubit, making them a versatile platform for sensing~\cite{rogers2014hybrid,chu2020perspective,kurizki2015quantum,clerk2020hybrid,lee2017topical, sinanan2024single}. The architecture has been proposed for a wide range of applications that include inertial-force sensing (such as gravimetry) and tests for macroscopic superposition in  levitated NV centers~\cite{scala2013matter,wan2016free,wan2016tolerance,yin2013large} and superconducting spheres~\cite{johnsson2016macroscopic,romero2012quantum,montenegro2025heisenberg,prat2017ultrasensitive,canturk2017quadrature}, magnetometry~\cite{hoang2016electron,neukirch2015multi,ji2025levitated,rani2025magnon}, electromechanical sensing in circuit-quantum electrodynamics (cQED)~\cite{pirkkalainen2015optomechanics,eichler2018longitudinal,bera2021large,potts2025longitudinal,billangeon2015circuit,johansson2014optomechanical} and dark matter detection~\cite{chen2024axion,braggio2025axion,dixit2021DM,agrawal2024DM,zhao2025DM}.

Squeezing the oscillator quadrature is a cornerstone of quantum sensing, enabling measurement sensitivities beyond classical or the standard quantum limit (SQL)~\cite{degen2017quantum,toth2014quantum}. This capability has opened the door to key experiments probing fundamental physics~\cite{carney2019tabletop, bose2025massive, rademacher2020quantum}. It is therefore desirable to seek a protocol that combines the high-fidelity qubit readout with the sensitivity enhancement provided by squeezing.  
{Although a straightforward application of squeezing amplifies the signal in displacement based sensing, we show that this enhancement does not carry over to qubit based measurement.}

In this Letter, we introduce a quantum sensing protocol for coupled qubit-oscillator systems that encodes the signal in a detectable geometrical phase. Importantly, this geometrical phase is proportional to the area enclosed by the oscillator motion in its phase space. 
Using the insight that squeezing is a geometrical operation that can amplify the oscillator path~\cite{burd2019quantum} and therefore the area, our protocol enables sensitivities beyond the SQL. Geometric phases in bosonic systems have been extensively studied~\cite{chaturvedi1987berry,pechal2012geometric,vacanti2012geometric}, applied to pulse calibration~\cite{eickbusch2022}, engineering entangling gates ~\cite{wang2002simulation, PhysRevA.100.043417,song2017continuous}, and even for generating entangled states for sensing~\cite{johnsson2020geometric}. 
{Here, we provide a framework that exploits the geometric nature of the phase, specifically, its proportionality to enclosed area to obtain enhanced sensitivity.}
We first apply our method to longitudinally coupled systems for force sensing such as gravimetry and then extend it to precision measurement of couplings and pulse calibration in  cQED systems that are dispersively coupled.  
Since the phase is accumulated along a closed loop in oscillator phase space, our protocol is independent of the oscillator’s initial state, allowing operation with high-temperature thermal states and compatibility with bosonic error-correcting codes such as GKP states~\cite{PhysRevA.64.012310, PRXQuantum.2.020101, liu2025}—opening pathways to fault-tolerant quantum sensing. Finally, we analyze the dynamics under Markovian noise affecting both qubit and oscillator, showing that the protocol retains its state-independence and robustness under qubit noise, underscoring its practicality for a broad range of next-generation quantum metrology applications.

\begin{figure}[!t]
    \centering
    \includegraphics[width=\columnwidth]{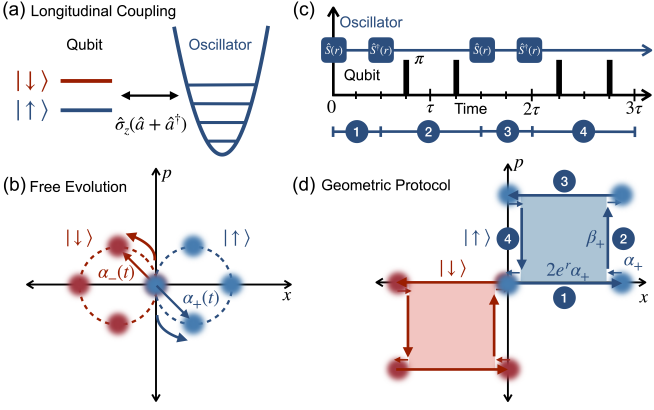}
 \caption{ We consider (a) longitudinally-coupled qubit-oscillator systems. {(b) Free evolution (solid) and effect of squeezing (dashed) for the coupled system in the oscillator phase space.} (c) Sequence of operations on the oscillator and qubit for the geometric protocol and (d) the corresponding phase space motion.}   
 \label{fig:Fig1}
\end{figure}

\textit{Sensing in longitudinally-coupled systems---} We first discuss detection of force by measuring the qubit in longitudinally coupled qubit-oscillator {(see Fig.~\ref{fig:Fig1}(a))} systems~\cite{scala2013matter,
wan2016free,
wan2016tolerance,
yin2013large,
johnsson2016macroscopic,
romero2012quantum,
montenegro2025heisenberg,
prat2017ultrasensitive,
canturk2017quadrature,
hoang2016electron,
neukirch2015multi,
ji2025levitated,
rani2025magnon,
pirkkalainen2015optomechanics,
eichler2018longitudinal,
bera2021large,
potts2025longitudinal,
billangeon2015circuit,
johansson2014optomechanical,
leibfried2003trap_ion,
haljan2005spin,
sorensen2000entanglement}. The Hamiltonian in the lab frame is given by 
\begin{align}
    \frac{H_L}{\hbar} = -\frac{\omega_q}{2} \sz + \omega \adag \a + \gamma(\a + \adag) \hat \sigma_z + \eta(\a + \adag) \;,
\end{align}
where $\omega_q$ is the qubit frequency, $\omega$ is the oscillator frequency, $\gamma$ is the longitudinal coupling strength, and $\eta = F/\sqrt{2m\omega\hbar}$ encodes an unknown weak force $F$ acting on the oscillator with mass $m$. Since $\sz$ commutes with $H_L$, we work in its eigen-basis with eigenvalues $s=\pm 1$ corresponding to eigenstates $\{\ket{\uparrow},\ket{\downarrow}\}$. For each qubit state $\ket{s}$, the time evolution in the frame rotating with qubit and oscillator is given by~\cite{supplemental} 
  \begin{align}
      \U_s(t) &= e^{is\frac{\phi(t)}{2}}\D(\alpha_s(t)) \;,
      \label{eqn:free}
  \end{align}
  {where the displacement $\D(\alpha) = \exp({\alpha \adag - \alpha^* \a})$}. 
  {The oscillator  traces a circular path with displacement $\alpha_s(t) = \alpha_s (1-e^{i\omega t})$ where $\alpha_s = (\eta + \gamma s)/\omega$, as shown in Fig.~\ref{fig:Fig1}(b). The qubit acquires the total geometrical phase $\Phi_s = \Im \int \alpha_s^*(t) d\alpha_s(t)$ proportional to the area enclosed by the oscillator motion. This phase is composed of a global part and an $s$-dependent detectable part $\Phi_{\text{det}}=s \phi(t)/2$, where $\phi(t)= 4\eta \gamma (\omega t- \sin\omega t)/\omega^{{2}}$.}
  Proposed sensing protocols~\cite{scala2013matter,wan2016free,wan2016tolerance,yin2013large,johnsson2016macroscopic,romero2012quantum,montenegro2025heisenberg,prat2017ultrasensitive} estimate the force by measuring the phase $\phi(t)$ using an interference experiment. We briefly summarize this free-evolution scheme in the following. 
The system is prepared in a product state $\ket{\psi(0)}=\ket{+} \ket{0}$, with the qubit in a $\sx$ eigenstate {$\ket{+}$} and the oscillator in the vacuum state $\ket{0}$. The system is evolved freely, after which $\sx$ is measured 
  \begin{align}
      \langle \sx \rangle = e^{-d(t)} \cos\phi(t) \;, 
      \label{eqn:sigmax}
  \end{align}
  where   $d(t) = (\alpha_+(t) - \alpha_-(t))^2/2={8\gamma^2}\sin^2(\omega t/2)/{\omega^2}$, quantifying the overlap of the displaced oscillator states. We calculate the quantum Fisher information (QFI) of the qubit with respect to the force $\eta$
  \begin{align}
      \mathcal{F}_\eta(t) = e^{-2d(t)}(\partial_\eta \phi(t))^2  \;,
      \label{eqn:QFI}
  \end{align}
the inverse of which is the minimum uncertainty in estimating the force. The QFI reaches its maximum at integer multiples of the oscillator period, $\tau = 2\pi/\omega$, where the two oscillator trajectories reconverge at the origin (Fig.~\ref{fig:Fig1}(b)) so that $d(n\tau)=0$, meaning the qubit becomes fully decoupled from the oscillator. Thus the information is contained solely in the phase $\phi(n\tau) = n \phi_0 $, where $\phi_0\equiv 8\pi \eta \gamma/\omega^2$.
Measuring $\sx$ at such times optimally extracts this information, confirmed by calculating the classical Fisher information  $\mathcal{F}^c_\eta(t) = (\partial_\eta \phi(t))^2\sin^2\phi(t)/(e^{2d(t)}-\cos^2\phi(t))$, which saturates the bound  $\mathcal F^c_\eta \leq \mathcal F_\eta$: $\mathcal F^c_\eta(n\tau) =  \mathcal F_\eta(n\tau) = 64 n^2 \pi^2 \gamma^2/\omega^4$.

{
We emphasize that a \emph{straightforward} application of squeezing to amplify displacement does not enhance sensitivity in this free-evolution scheme.
Conjugating the free evolution in Eq.~\ref{eqn:free} with squeezing, $\hat{S}^\dagger(r)\, \hat U_s(t)\hat{S}(r)$, 
transforms $\alpha_s(t) \rightarrow \alpha_s (e^r (1-\cos\omega t) -i e^{-r}\sin\omega t)$,
deforming the circular trajectory into an ellipse (Fig.~\ref{fig:Fig1}(b)). The transformation preserves the enclosed area, leaving the detectable phase $\phi(t)$ unchanged~\cite{supplemental}.
This is a general consequence of symplectic invariance: the geometric phase is $\Phi_s = \frac{1}{2}\oint \bm{r}^T J d\bm{r}$, where $\bm{r} = (\langle \x \rangle_{\alpha_s}, \langle \p \rangle_{\alpha_s})^T$ and $J=i \hat \sigma_y$.
Any Gaussian transformation $\bm r^\prime = M \bm r + \bm q$ is symplectic ($M^T J M = J$) to preserve the operator commutation relations~\cite{weedbrook2012gaussian,liu2025}. For time-independent $M,\bm q$, this immediately implies $\Phi^\prime = \Phi$ for a closed loop. 
As a result, widely used squeezing $M=\text{diag}(e^r,e^{-r})$ applied anywhere and for any duration, preserves the area which results in the SQL limited sensitivity.}

\textit{Geometric Protocol---} 
{
We now present a sensing protocol that enables beyond SQL sensitivity of force by enhancing the geometrical phase. A general protocol must ensure that (i) the oscillator loop is closed $\oint dr = 0$ amounting to only a phase kickback on the qubit, (ii) the $s$-dependent detectable part of the that phase contains the force $\eta$; and (iii) symplectic invariance is broken to enable amplification through squeezing. We provide general conditions in the supplemental~\cite{supplemental} and here, consider a minimal example that uses squeezing on the oscillator and $\pi$-pulses on the qubit, both readily available across many platforms.  The scheme is illustrated in Fig.~\ref{fig:Fig1}(c),(d), and we highlight the key steps in the following.
}

In step 1, we conjugate the free evolution with a squeezing operation to achieve amplification of the signal,
\begin{align}
\U_{1,s} &= e^{i\frac{\phi_0}{4}s} \D(2e^r \alpha_s) \;.
\end{align}
In step 2, we apply two $\pi$-pulses to the qubit at times shown in Fig~\ref{fig:Fig1}(c), 
implementing displacements 
\begin{align}
\U_{2,s}=\D(\alpha_s)\D({i}\beta_s)\Ddag(\alpha_s)\;,   
\end{align}
where $\beta_s= 4 \gamma s/\omega$ is the displacement along the $p$ axis.
The evolution in the oscillator phase space is depicted in Fig.~\ref{fig:Fig1}(d). 
To complete the loop, we repeat the first two steps, time shifted by $\tau/2$, obtaining 
\begin{align}
\begin{split}
\U_{3,s} &= e^{i\frac{\phi_0}{4}s} \Ddag(2e^r \alpha_s), \\
\U_{4,s} &= \D^\dag(\alpha_s)\D^\dag({i}\beta_s)\D(\alpha_s) \;,
\end{split}
\end{align}
Taken together, the total evolution traces a rectangular path in the phase space given by 
\begin{align}
\begin{split}
        \U_s &= \U_{4,s} \U_{3,s} \U_{2,s} \U_{1,s} =e^{i\frac{\phi_T}{2}} \hat{I} \;,
    \end{split}
    \end{align}
where the total phase $\phi_T = \phi_0 s + 4 A_s$ and 
\begin{align}
\label{eq:As}
     A_s = 2\alpha_s(e^r-1)\beta_s = 8(e^r-1) s\gamma \eta/\omega^2 {+ \text{global}}
\end{align}
is the additional geometrical contribution, equal to the area enclosed by the oscillator motion marked by the shaded region in Fig.~\ref{fig:Fig1}(d). 
The $\pi$-pulses ensure that the oscillator traces a closed loop and importantly engineers the conditional displacement $\beta_s$. When $\beta_s$ is combined with the force $\eta$ in $\alpha_s$, as in Eq.~\ref{eq:As}, it guarantees that the force $\eta$ is contained in the $s-$dependent part of the geometrical phase, hence detectable. The oscillator state is shown at specific snapshots in time. The path will look different when viewed at different times; however, the area remains the same, highlighting its geometric nature. The total evolution decouples the qubit and the oscillator, amounting to only a phase kickback on the qubit and identity operation on the oscillator, indicating that our protocol is independent of the initial oscillator state. This has the benefit of using higher temperature thermal states that are easy to prepare or logical states such as GKP that can aid in error corrected sensing.

 \begin{figure}
    \centering
    \includegraphics[width=\columnwidth]{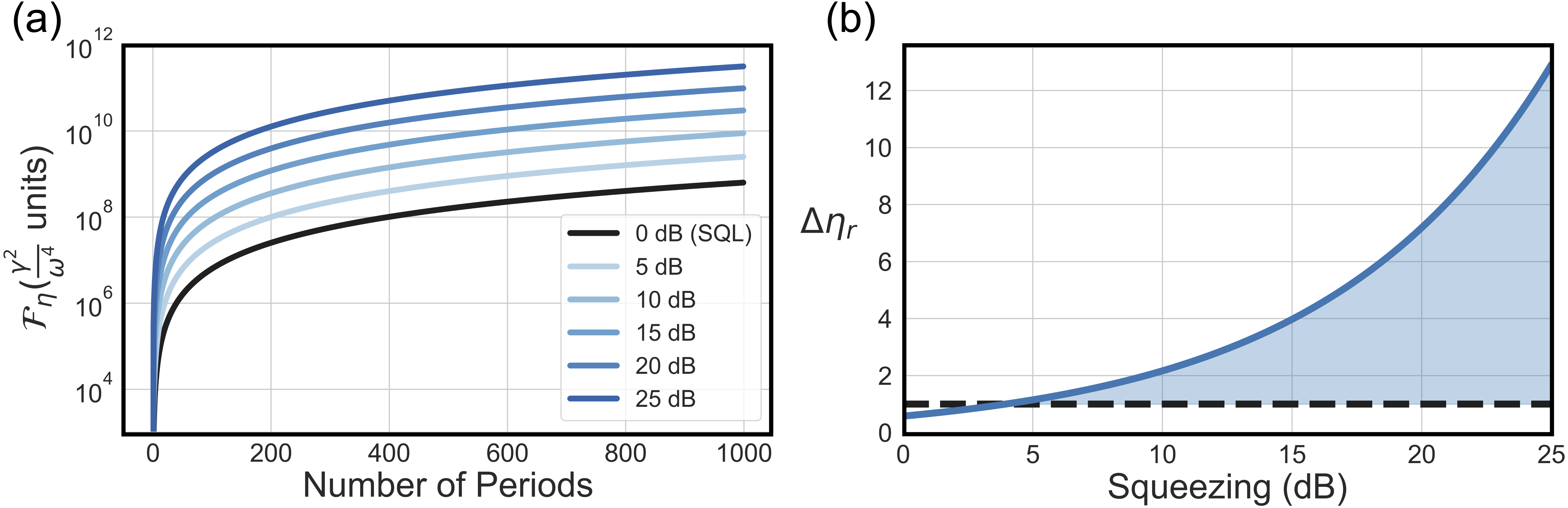}
    \caption{(a) Quantum Fisher information of the qubit to sense the force $\eta$ as a function of time. (b) The relative sensitivity of geometric protocol in comparison to standard free evolution scheme as a function of squeezing. Dashed line shows $\Delta\eta_r = 1$ as a reference.}
    \label{fig:Fig2}
\end{figure}

We therefore start in an arbitrary oscillator state with the qubit in $\ket{+}$ superposition state,  evolve under $U_s$, and perform an interference experiment similar to the free-evolution scheme by measuring $\sigma_x$ of the qubit to extract the phase.
Using Eq.~\ref{eqn:QFI}, we calculate the QFI for the geometric protocol 
\begin{align}
\label{eq:QFIGeo}
    \mathcal F^{\text{g}}_\eta &= 64 n^2  (\pi + 4 (e^r -1))^2 \gamma^2/\omega^4\;,
\end{align}
where $n$ is the number of loops completed in the phase space. It is crucially dependent on the area enclosed in the phase space and therefore increases with the squeezing parameter $r$, thus confirming beyond-SQL scaling, as plotted in Fig~\ref{fig:Fig2}(a). For $r=0$, the geometric part vanishes, leaving the standard quantum limited measurement shown with the black line. 
To evaluate improvement in sensitivity with respect to the free-evolution protocol, we account for the longer duration of the geometric loop, which takes $T_g = 3\tau$, three times that of the free-evolution period. 
We compare the sensitivity of the geometric protocol $\Delta \eta_g = 1/\sqrt{\mathcal F^g_\eta/(nT_g)}$, with that of the free-evolution scheme $\Delta \eta_0 = 1/\sqrt{\mathcal F_\eta(n\tau)/(n \tau)}$. Figure~\ref{fig:Fig2}(b) shows the relative sensitivity $\Delta \eta_r = \Delta \eta_0/\Delta \eta_g$ as a function of squeezing, where $\Delta \eta_r>1$ indicates the gain obtained from using the geometric scheme. At low squeezing, the free-evolution scheme yields better sensitivity due to the shorter cycle time. However, beyond $\approx 4~\text{dB}$ of squeezing, the geometric contribution outperforms, highlighted by the shaded region. State-of-art platforms already achieve $\sim 15~\text{dB}$ of squeezing \cite{cai2025squeezing} corresponding to $4$ fold improvement over SQL limited free evolution protocol.

\begin{figure}[!t]
    \centering
    \includegraphics[width=\columnwidth]{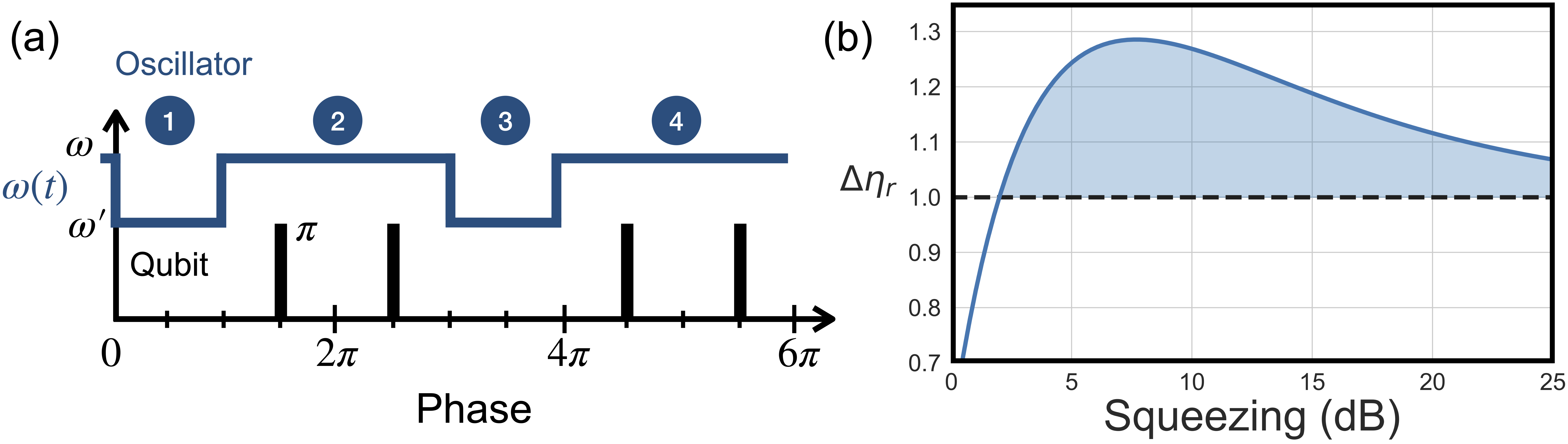}
    \caption{(a) Sequence of operations for the geometric method with phase $\theta = \int dt \omega(t)$, where the squeezing is achieved by suddenly changing the oscillator frequency $\omega(t)$ from $\omega$ to $\omega^\prime$. (b) Relative sensitivity of geometric protocol in comparison to standard free evolution with  $\omega^\prime$ frequency as a function of squeezing.}
    \label{fig:Fig3}
\end{figure}
Next, we apply our method to the case where squeezing is explicitly achieved via a sudden change in the oscillator frequency~\cite{PhysRevLett.127.183602} from $\omega$ to $\omega^\prime$, which implements squeezing with strength $r = \ln(\omega/\omega^\prime)/2 $, with protocol illustrated in Fig.~\ref{fig:Fig3}(a). Evolving the oscillator at a lower frequency $\omega^\prime$, rescales both the displacement amplitude  ${\alpha_s^\prime = \alpha_s e^{2r}}$, and the phase collected per evolution $\phi_0^\prime = \phi_0 e^{4r}$. However, this exponential factor is trivial as it arises solely from evolving the oscillator at a lower frequency and not from squeezing. 
Steps 2 and 4 of our protocol remain unchanged, while steps 1 and 3 are rescaled as $\U_{1,s} =e^{is\phi^\prime_0/4}\D(2\alpha_s e^{3r})$ and $\U_{3,s}=e^{is\phi^\prime_0/4}\D^\dag(2\alpha_s e^{3r})$~\cite{supplemental}. The total relative phase is now $\phi_T =  \phi^\prime_0 s + 4 A_s$, where the additional geometric contribution is $A_s =8 (e^{3r}-1) {s} \eta \gamma / \omega^2$. The factor $e^{3r}$ arises purely from the oscillator trajectory in phase space and contributes to the total phase, enhancing the sensing performance. This is confirmed by the corresponding quantum Fisher information $\mathcal F_\eta = 64n^2  (e^{4r}\pi + 4(e^{3r}-1))^2\gamma^2/\omega^4$. The time taken for the total loop is now $T_g = 2 \times 2\pi/\omega  + 2\pi/\omega^\prime$ as the oscillator spends one period with $\omega^\prime$ frequency and two periods in $\omega$. Although this is longer than the freely evolving the oscillator under lower frequency $\omega^\prime$, the additional $e^{3r}$ enhancement in phase improves the sensitivity. The relative sensitivity of the geometric protocol compared to free-evolution under $\omega^\prime$ is shown in Fig.~\ref{fig:Fig3}(b), demonstrating up to $\sim 30\%$ improvement for current experimentally achievable squeezing~\cite{PhysRevLett.127.183602}.

\textit{Sensing in dispersively-coupled systems---}  
We now extend our method to dispersively coupled qubit–oscillator systems, which are widely realized in standard superconducting quantum devices using circuit quantum electrodynamics (cQED) architectures~\cite{blais2004cavity, krantz2019cqed, clerk2020hybrid}, where high-fidelity qubit readout is routinely available~\cite{sunada2022readout, kim2025ultracoherent}. In the dispersive regime, the effective Hamiltonian is given by
\begin{align}
    H_D/\hbar = \frac{\chi}{2} \adag \a \sz + i (\epsilon(t) \adag - \epsilon^*(t) \a) \;,
\end{align}
where $\chi$ is the dispersive shift and $\epsilon(t)$ is the driving pulse on the {resonator}.  The pulse amplitude {$\varepsilon_0 = \max(\epsilon(t))$} is generally unknown apriori and was determined using a geometrical phase~\cite{eickbusch2022} which is SQL limited. We demonstrate that our protocol can be used to enhance this sensitivity beyond the SQL. Additionally, the same method can measure dispersive shift $\chi$ surpassing SQL.

\begin{figure}[!t]
    \centering
    \includegraphics[width=1\columnwidth]{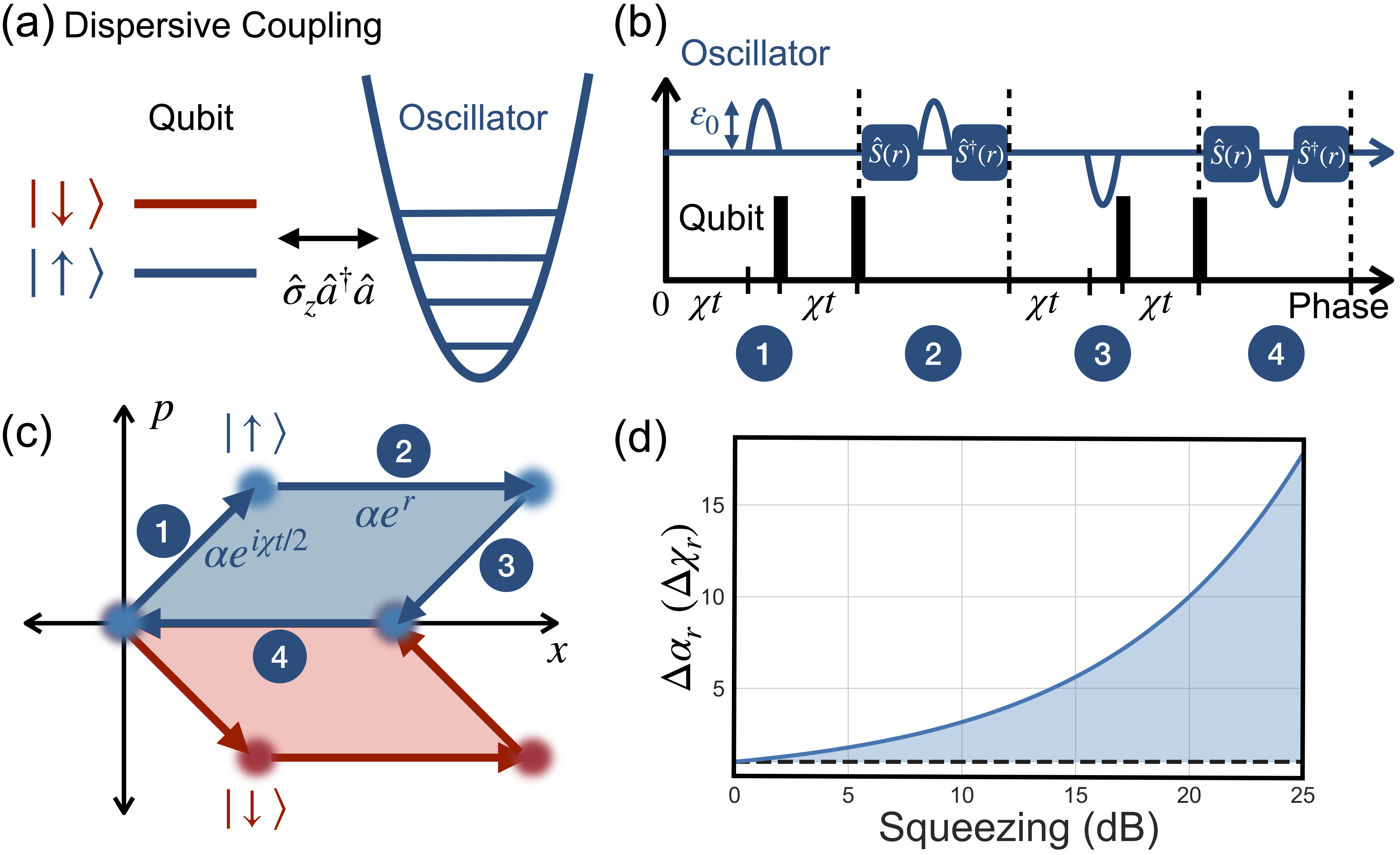}
    \caption{(a) We consider dispersively coupled qubit-oscillator systems. (b) Sequence of operations on the oscillator and qubit for the geometric protocol and (c) the corresponding phase space evolution. (d) Relative sensitivity in comparison to SQL for estimating either the displacement $\alpha$ or the dispersive coupling $\chi$ as a function of squeezing.}
    \label{fig:Fig4}
\end{figure}

The procedure is illustrated in Fig.~\ref{fig:Fig4}(b),(c), where we steer the system around a loop~\cite{supplemental}. A related protocol for large spins, but without squeezing, was explored in Ref.~\cite{johnsson2020geometric}. 
{In step 1, the system evolves freely for a time $t$ under
$\R = e^{-is\frac{\chi t}{2}\adag \a}$. The cavity is then strongly driven
($\varepsilon_0 \gg \chi$) to generate a displacement $\alpha = \int \epsilon(t) dt$
along $x$, followed by a reversal of the free evolution,
\begin{align}
    \U_1 = \R^\dag \D(\alpha)\R = \D(\alpha e^{is\frac{\chi t}{2}})\;,
\end{align}
where $\R^\dag$ is implemented by two $\pi$ pulses that flip $\sz$, the first immediately after the drive and the second an interval $t$ later~\footnote{In the strong driving limit, the displacement is treated as instantaneous since the drive duration is negligible compared with the free evolution timescale: $t_{\text{drive}} \simeq \alpha/\varepsilon_0 \ll 1/\chi$}.}
Subsequently, we conjugate the drive with squeezing to obtain amplification of the signal
\begin{align}
    \U_2 = \S^\dag(r) \D(\alpha)\S(r) = \D(\alpha e^r) \;.
\end{align} 
Finally, we reverse the first and second unitary by applying the resonator drive with a phase shift of $\pi$, thereby closing the loop $\U = \U_2^\dag\U_1^\dag\U_2\U_1$, as depicted in Fig.~\ref{fig:Fig4}(b),(c). At the end of the process, the oscillator and qubit decouple, yielding $\U = e^{i s \phi/2} \hat{I}$,  with the relative phase on the qubit $\phi = 4 A$, where $A = \alpha^2 e^r \sin (\chi t/2)$ is the area enclosed in the parallelogram traversed by the oscillator in the phase space~\cite{supplemental}. This phase is measured by a similar interference experiment by starting in an arbitrary oscillator state and qubit $\ket{+}$ state. It is used to estimate the displacement to obtain the pulse amplitude or the dispersive coupling given that the other is known. The corresponding QFI using  Eq.~\ref{eqn:QFI} after traversing $n$ loops in the phase space is 
\begin{align}
\begin{split}
     \mathcal F_\alpha &= (\partial_\alpha \phi)^2 = 64 n^2 \alpha^2 e^{2r} \sin^2(\chi t/2) \;,\\
 \mathcal F_\chi &= (\partial_\chi \phi)^2 = 4 n^2 t^2 \alpha^4 e^{2r} \cos^2(\chi t/2) 
 \;.
 \end{split}
 \end{align}
Figure~\ref{fig:Fig4}(d) shows the relative enhancement in sensitivity $\Delta \alpha_r, \Delta \chi_r = e^r$, as compared to SQL ($r=0$). Given experimentally demonstrated squeezing of $15~\text{dB}$~\cite{cai2025squeezing}, the method obtains $\sim 5$ times improvement over SQL for both pulse calibration and dispersive shift estimation.

\begin{figure*}[!t]
    \centering 
\includegraphics[width=2\columnwidth]{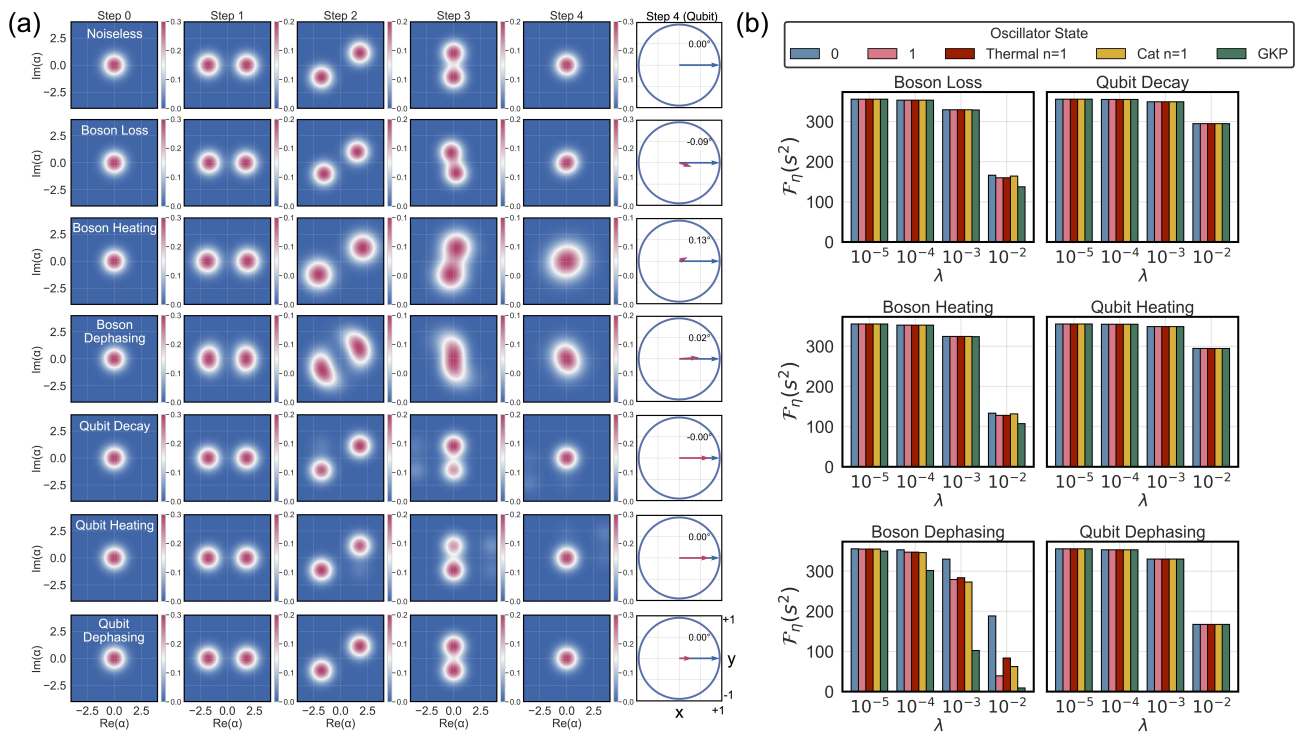}
    \caption{Effect of noise for inertial force sensing. We consider typical parameters where $\omega=1 $, qubit-oscillator coupling $\gamma = 0.2$, with $10~\text{dB}$ of squeezing, the force $\eta=10^{-5}$ and the fock space cutoff $N=200$. (a)  Wigner functions for the joint state at each step of the geometric protocol for all noise channels with strength $\lambda=0.05$. The last column is the {final} qubit state projected to the $x-y$ plane (marked by red arrow), with printed phase $\phi$ relative to the noiseless case (shown with blue arrow). For visual clarity the vectors are drawn with $100 \times$relative angle. (b) Numerical calculation of QFI for different initial states after 1 period of evolution for each noise channel. {The error rates $\lambda$ are chosen to reflect realistic experimental conditions across different physical systems~\cite{supplemental}.}
    }
    \label{fig:Fig5}
\end{figure*}

\textit{Robustness  ---} 
Preserving sensitivity under noise is crucial as realistic systems inevitably suffer from dissipation and decoherence. We show that the geometric nature of our protocol {makes it resilient} to qubit noise while retaining its independence of the oscillator’s initial state. 
We numerically simulated the system under Markovian noise using a Lindblad master equation acting on both the oscillator and the qubit~\cite{hastrup2021unconditional}. 
As a representative case, we analyze the inertial sensing protocol by interleaving the unitary steps $\mathcal U_i(\bullet) = \U_i(\bullet)\hat{U}_i^\dag$, $i\in\{1,\ldots,4\}$ with noise channel $\mathcal E_t = e^{\mathcal L t}$, 
where the Lindblad term is  $\partial_t \rho = \mathcal L \rho = \hat L \rho \hat L^\dag - \frac{1}{2}(\hat L^\dag \hat L \rho + \rho \hat L^\dag \hat L)$ and $\rho$ is the joint system state. We consider {a comprehensive list of} six {elementary} noise processes: (i) boson loss $\hat L = \sqrt{\lambda} \a$, (ii) boson heating $\hat L = \sqrt{\lambda} \adag$, (iii) boson dephasing $\hat L = \sqrt{\lambda} \adag \a$, (iv) qubit decay $\hat L = \sqrt{\lambda} \sigma^-$, (v) qubit heating $\hat L = \sqrt{\lambda} \sigma^+$, and (vi) qubit dephasing $\hat L = \sqrt{\lambda} \sz$, where $\lambda$ is the rate of noise. The noisy evolution is given by 
\begin{align}
    \mathcal E_T =\mathcal E_{\tau}  \circ \mathcal U_4 \circ \mathcal E_{\tau/2} \circ \mathcal U_3 \circ \mathcal E_{\tau}   \circ \mathcal U_2 \circ \mathcal E_{\tau/2} \circ \mathcal U_1 \;,
\end{align}
where $\circ$ represents the composition of quantum channels and each dissipative channel has the same duration as the corresponding unitary step.

For each noise channel, the Wigner function is shown at each step of the evolution in Fig.~\ref{fig:Fig5}(a) . 
The protocol demonstrates much stronger robustness to qubit noise than to bosonic noise. Bosonic noise perturbs the oscillator phase-space trajectory, distorting the area which corrupts the signal. The last column shows qubit state projected to the $x-y$ {Bloch} plane, where the phase $\phi_T$ encoding the signal {(red arrow)} is shown relative to the noiseless phase {(blue arrow)}, highlighting the deviation under bosonic noise. By contrast, qubit decay (excitation) merely redistributes the population between $\ket{0}$ and $\ket{1}$, leaving the oscillator’s trajectory largely intact, while qubit dephasing commutes with the unitaries, producing an overall decay factor, we see that this leaves the angle and thus the signal largely intact, indicating robustness.

Finally, we examine the protocol's insensitivity to different initial oscillator states by considering five representative states: the vacuum state, coherent state $\alpha=1$, thermal state $n=1$, cat state $n=1$, as well as the GKP logical state. Figure~\ref{fig:Fig5}(b) shows the QFI for each noise channel.  Consistent with the Wigner function analysis, the QFI for qubit channels is much higher than their bosonic counterpart. Importantly, the protocol retains its independence over the initial oscillator states near perfectly under qubit noise.

In summary, we introduced a quantum sensing protocol for qubit-oscillator systems that harnesses the geometrical phase to obtain sensitivities beyond the SQL. The method is resilient to qubit noise which can be a dominant source of error, offering practical advantages in real world applications. This work reveals another interesting aspect of the much-studied geometrical phases and their natural intersection with beyond-SQL sensing, opening new possibilities in quantum metrology.

\begin{acknowledgments}
We thank Daniel Carney, Giacomo Marocco and Adrian Lupascu for fruitful discussions. This work is supported by a collaboration between the US DOE and other agencies. NS acknowledges support from the ``Embedding Quantum Computing into Many-body Frameworks for Strongly Correlated Molecular and Materials Systems'' project, funded by the US DOE, Office of Science, Office of Basic Energy Sciences, Division of Chemical Sciences, Geosciences, and Biosciences. 
WAdJ is supported by the US DOE, Office of Science, National Quantum Information Science Research Centers, Quantum Systems Accelerator at Lawrence Berkeley National Laboratory under Contract No.~DE-AC02-05CH11231. 
ZW and DV were partially supported by NASA(80ARC020D0010), and ZW acknowledges internal IR\&D funding. TR and DV acknowledge support from the US DOE, Office of Science, National Quantum Information Science Research Centers, Superconducting Quantum Materials and Systems Center (SQMS), under Contract No.~89243024CSC000002.
\end{acknowledgments}

\bibliography{Sensing}
\end{document}


\title{Supplemental Material on ``Quantum Sensing Using Geometric Phase in Qubit-Oscillator Systems"}
\author{Nishchay  Suri}\thanks{surinishchay@gmail.com}
\affiliation{Applied Mathematics and Computational Research Division, Lawrence Berkeley National Laboratory, Berkeley, California 94720, USA}
\author{Zhihui Wang}
\affiliation{Quantum Artificial Intelligence Laboratory, NASA Ames Research Center, Moffett Field, CA 94035, USA}
\affiliation{USRA Research Institute for Advanced Computer Science, Mountain View, California 94043, USA}
\author{Tanay Roy}
\affiliation{Superconducting Quantum Materials and Systems (SQMS) Center),
Fermi National Accelerator Laboratory (FNAL), Batavia, IL 60510, USA}
\author{Davide Venturelli}
\affiliation{Quantum Artificial Intelligence Laboratory, NASA Ames Research Center, Moffett Field, CA 94035, USA}
\affiliation{USRA Research Institute for Advanced Computer Science, Mountain View, California 94043, USA}
\author{Wibe Albert de Jong}
\affiliation{Applied Mathematics and Computational Research Division, Lawrence Berkeley National Laboratory, Berkeley, California 94720, USA}
\maketitle

\section{Longitudinally coupled Qubit-Oscillator Systems}
\subsection{Free Evolution}
The Hamiltonian of the coupled qubit oscillator system in the lab frame is given by 
\begin{align}
    \frac{H_L}{\hbar} = -\frac{\omega_q}{2} \sz + \omega \a^\dag \a + \gamma (\a + \a^\dag) \hat\sigma_z + \eta (\a + \a^\dag)\;,
\end{align}
{where} $\eta = F \frac{1}{\sqrt{{2m \omega \hbar}}}$ and $F$ is a general force ($F=mg$ for gravity). We do our entire analysis in the rotating frame $U_0(t) = e^{-i (\frac{\omega_q}{2} \sz - \omega \a^\dag \a)t}$ where the Hamiltonian now is given by 
\begin{align}
    H(t) = (\a e^{-i\omega t} + \adag e^{i\omega t}) (\eta + \gamma \sz) \;,
    \label{eq:H(t)}
\end{align}
where $H$ is in units of $\hbar$. Notice that this is a time dependent periodic Hamiltonian, however we can write its exact solution using Floquet framework. The harmonics of the system are given by $H_0 = 0$, $H_{+1} = \adag (\eta + \gamma \sz)$ and $H_{-1} = \a (\eta + \gamma \sz)$. Using Floquet theory, the total evolution of the system is given by
\begin{align}
    U(t,t_0) = e^{-iK(t)} e^{-i H_F(t-t_0)} e^{+i K(t_0)} \;,
\end{align}
where the Floquet Hamiltonian is  
\begin{align}
    H_F  = - \frac{1}{\omega} (\eta + \gamma \sz)^2
\end{align}
and the micromotion operator
\begin{align}
    K(t) =  \frac{1}{i\omega} (\adag e^{i\omega t} - \a e^{-i\omega t}) (\eta+\gamma \sz) \;.
\end{align}
The Floquet Hamiltonian only operates on the qubit, and as a result it commutes with micromotion operator $[H_F, K(t)] = 0$. Moreover, everything commutes with $\sz$, we can work in its eigenbasis $\ket{s} \in \{ \ket{\uparrow}, \ket{\downarrow}\}$, corresponding to eigenvalues $s = \pm 1$. Defining $\alpha_s = (\eta + \gamma s)/{\omega}$, the total evolution becomes
\begin{align}
\begin{split}
        U_s(t,t_0) &= e^{i\frac{2\eta \gamma s}{\omega}(t-t_0)}\D^\dag(\alpha_s e^{i\omega t}) \D(\alpha_s e^{i\omega t_0}) \;.
        \label{eq:free_evolve}
\end{split}
\end{align}
Starting at $t_0=0$ the evolution is
\begin{align}
    U_s(t,0) = e^{is\phi(t)/2} \D(\alpha_s(t))\;,
\end{align}
where $\phi(t) = 4\eta\gamma (\omega t-\sin\omega t)/\omega^{{2}}$ and $\alpha_s(t) = \alpha_s (1-e^{i\omega t})$.

{
\textit{Evolution from infinitesimal displacements}--- 
We now explicitly show that the phase $\phi(t)$ is geometric by considering an alternate derivation from a series of infinitesimal displacements~\cite{leibfried2003experimental}. We use the identity $\D(\Delta \alpha_2)\D(\Delta \alpha_1) = \D(\Delta\alpha_1 + \Delta \alpha_2)e^{i\Im(\Delta\alpha_2 \Delta\alpha_1^*)}$.  Then for an arbitrary series of displacements, we have
\begin{align}
    \D &= \D(\Delta \alpha_N) \dots \D(\Delta \alpha_1) \\
    &= \D(\sum_j \Delta \alpha_j) e^{i \Im (\sum_{j=2}^N \Delta\alpha_j (\sum_{k=1}^{j-1}\Delta\alpha_k)^*)}
\end{align}
To go to the continuum limit, we replace $\Delta\alpha_j = d\alpha$, then $\sum_{k=1}^{j-1}\Delta\alpha^*_k = \alpha^*(t_{j-1})$. We then obtain
\begin{align}
    \D &= \D(\int \frac{d\alpha}{dt} dt) e^{i\Phi} \;, &
    \Phi &= \Im \int \alpha^*(t)  d\alpha(t) \;,
\end{align}
where $\Phi$ is the geometrical phase proportional to the area enclosed by the series of oscillator displacements. In our system, an infinitesimal displacement generated by the Hamiltonian in Eq.~\ref{eq:H(t)} is $U(\Delta t) = e^{-i H(t) \Delta t} = \D(\Delta \alpha)= e^{\Delta \alpha \adag - \Delta \alpha^* \a}$, where $\Delta \alpha = -i(\eta+\gamma s)e^{i\omega t}\Delta t$. Taking the continuum limit we then have 
\begin{align}
   d\alpha &= -i(\eta+\gamma s) e^{i\omega t} dt \;, & \alpha_s(t) &=  \alpha_s (1-e^{i\omega t}) \;,
\end{align}
which is exactly the displacement in our derivation. The total evolution is given by 
\begin{align}
    U(t,0) &= e^{i\Phi_s} \D(\alpha_s(t)) \;, &
    \Phi_s &= \alpha_s^2 (\omega t - \sin\omega t)\;.
    \label{eqn:Geometric}
\end{align}
Throwing away the global part, we obtain the $s$-dependent geometric phase $\Phi_{\text{det}}(t) = s \phi(t)/2$, where $\phi(t) = 4\eta\gamma (\omega t - \sin\omega t)/\omega^2$. For a closed loop $\Phi_s = 2 \pi \alpha_s^2$, which is the area of the circle traversed by the oscillator in Fig.~1(b). 
}


\textit{Free evolution protocol---} We start with the initial state 
\begin{align}
    \ket{\psi(0)} &= \frac{1}{\sqrt{2}}(\ket{\uparrow} + \ket{\downarrow} ) \otimes \ket{0} \;, 
\end{align}
and evolve the system under the free evolution Hamiltonian given above and obtain
\begin{align}
    \ket{\psi(t)} &= \frac{1}{\sqrt{2}} (\ket{+} U_+(t) \ket{0} + \ket{-}U_-(t)\ket{0}) \;,
\end{align}
where the reduced state of the qubit is 
\begin{align}
    \rho_q(t) &=\frac{1}{2}
    \begin{bmatrix}
        1 & e^{-i\phi(t)} \bra{\alpha_-(t)}\ket{\alpha_+(t)}\\
        e^{i\phi(t)} \bra{\alpha_+(t)}\ket{\alpha_-(t)} & 1 
    \end{bmatrix} \;.
\end{align}
Measuring $\sx$ of the qubit state in fact directly measures this overlap as 
\begin{align}
    \langle \sx\rangle = \frac{1}{2} (e^{-i\phi(t)} \bra{\alpha_-(t)}\ket{\alpha_+(t)} + e^{i\phi(t)} \bra{\alpha_+(t)}\ket{\alpha_-(t)}  ) = \cos(\phi(t)) e^{-d(t)} \;,
\end{align}
where $d(t) = 2(\alpha_- - \alpha_+)^2 \sin^2(\omega t/2)$ quantifies the overlap of the oscillator states.

\textit{Fisher Information---} 
{
We now derive the optimality of measuring $\sx$ from the information perspective. 
For a general partial quantum state $\rho_p=\sum_n \lambda_n \ket{\lambda_n}\bra{\lambda_n}$, the QFI is defined as~\cite{Liu_2020}
\begin{align}
    \mathcal F_\theta = 2 \sum_{m,n} \frac{|\langle \lambda_m|\partial_\theta \rho_p |\lambda_n \rangle|^2}{\lambda_m+\lambda_n} \;, \lambda_n+\lambda_m \neq 0 \;,
        \label{eq:partial-quantum-fisher-information}
\end{align}
which measures the maximum sensitivity of state $\rho_p$ to changes in $\theta$.
Classical Fisher Information (CFI) measures the sensitivity in parameter $\theta$ upon performing a measurement of operator $\Gamma_x$ leading to probabilities $p(x, \theta)$
\begin{align}
    \mathcal F^c_\theta = \sum_x p(x,\theta) \big[\partial_\theta \ln p(x,\theta)\big]^2\;.
\end{align}
The CFI achievable over all possible quantum measurements is bounded by QFI: $\Delta \theta^2 \geq \frac{1}{\nu\mathcal F^c_\theta} \geq \frac{1}{\nu\mathcal F_\theta}$, where $\nu$ is the number of experiments. We now calculate Fisher information for the qubit with respect to $\eta$. 
Through the reduced density matrix, $\rho_q(t)$,} we obtain its QFI
\begin{align}
    \mathcal F_\eta (t) &=  ({\partial_\eta\phi(t)})^2 e^{-2d(t)} \;.
    \label{eqn:qfi}
\end{align}
To check if our measurement of $\sx$ is optimal we calculate the classical Fisher information given by
\begin{align}
  \mathcal  F^c_\eta(t) = \frac{\sin^2(\phi(t))}{e^{2d(t)}-\cos^2(\phi(t))} (\partial_\eta \phi(t))^2 \;.
      \label{eqn:cfi}
\end{align}
Since $d(t)$ is a positive function, the QFI is maximized when $d(t)=0$. Moreover, the classical Fisher information saturates the bound when $d(t)=0$ which is when it extracts the maximum information contained in the system. This happens at integer time periods of the oscillator when the qubit and oscillator are decoupled and the damping factor vanishes at $\omega t_n = 2 n \pi$.
Putting in all the values we get for a general time
\begin{align}
  \mathcal F_\eta (t) &= 16 \frac{\gamma^2}{\omega^4} (\omega t-\sin\omega t)^2 e^{-16 \frac{\gamma^2}{\omega^2} \sin^2(\omega t/2)} \;.
\end{align}
and at integer times 
\begin{align}
    \mathcal F_\eta(t_n) = 64 n^2 \frac{\gamma^2\pi^2}{\omega^4} \;.
\end{align}

\textit{Squeezing---} 
{
We now conjugate the free evolution by squeezing $U =\S^\dag(r) U(\Delta t_N) \dots U(\Delta t_1) \S(r)$. We can use the identity $S^\dag(r) S(r) = I$ everywhere and define the infinitesimal displacement $U(\Delta t) = e^{-i \S^\dag(r) H(t)S(r) \Delta t}$, where 
\begin{align}
    \S^\dag(r) H(t) \S(r) &= (\eta+\gamma s) \bigg((
    \a\cosh r - \adag \sinh r) e^{-i\omega t} + (\adag \cosh r - \a \sinh r) e^{i\omega t} \bigg) 
\end{align}  
Again comparing with the infinitesimal displacement operator and taking the continuum limit we have    $ d\alpha = -i(\eta + \gamma s) ( e^{i\omega t }\cosh r - e^{-i\omega t} \sinh r)  dt $, integrating it we obtain
\begin{align}
    \alpha(t) = \alpha_s \bigg(  e^r(1-\cos\omega t) - ie^{-r} \sin\omega t \bigg)
\end{align}
This is the equation of an ellipse. Parameterizing it into real and imaginary parts, such that $\alpha = \frac{1}{\sqrt{2}} (x + i p)$, then $x=\sqrt{2} \alpha_s e^r (1-\cos\omega t)$ and $p = -\sqrt{2}\alpha_s e^{-r} \sin\omega t$, where $\Phi_s=\frac{1}{2}\int xdp-pdx=\alpha_s^2 (\omega t-\sin\omega t)$ the same as earlier without squeezing. More explicitly $x,p$ satisfy $(1-x/a)^2 + p^2/b^2 = 1$, where $a=\sqrt{2}\alpha_s e^r, b= \sqrt{2}\alpha_s e^{-r}$. One axis is elongated while the other shrinks keeping the area $\Phi_s=\pi ab = 2\pi \alpha_s^2$ the same. We now explain this using a more general framework below.}

{\subsection{Beyond SQL sensing with geometric phase} 
\textit{Area Invariance---} The quadrature operators $\hat{\bm{r}}= (\x, \p)^T$ satisfy the commutation relation $[\hat r_i,\hat r_j] = i J_{ij}$, where $J=\begin{bmatrix}
    0 & 1\\
    -1 & 0
\end{bmatrix}$. The most general Gaussian transformation on $\hat{\bm{r}}$ is an affine transformation $\hat{\bm{r}}^\prime = M \hat{\bm r} + \bm{q}$. Since the commutator is always preserved, the matrix $M$ is symplectic or $M^T J M = J$~\cite{weedbrook2012gaussian,liu2025}. For a coherent state $\ket{\alpha}: \a \ket{\alpha}=\alpha \ket{\alpha}$ which we parametrize as $\alpha = (x+ip)/\sqrt{2}$, where $x = \langle \x \rangle_\alpha$ and $p = \langle \p \rangle_\alpha$, the geometric phase can be written as
\begin{align}
    \Phi = \frac{1}{2}\oint \bm{r}^T J d\bm r \;,
\end{align}
where $\bm r = (x,p)^T$. The same $J$ that appears in the commutation relation, appears in the geometric area calculation as it naturally arises out of successive displacement commutators in time. For time-independent Gaussian transformation $(M,\bm q)$, the area remains invariant as
$  
    \Phi^\prime = \frac{1}{2}\oint \bm{r}^{\prime T} Jd \bm{r}^\prime  = \frac{1}{2} \oint ({\bm r}^T M^T  + \bm{q}^T) J M d{\bm r} = \Phi,
$
which follows immediately from the symplectic nature of $M: M^T J M = J$ and the second term vanishes for a closed loop. Therefore, any time-independent Gaussian operation such as widely used squeezing and anti-squeezing where $M = \text{diag} (e^r, e^{-r})$ and $\bm q = 0$, placed anywhere in the loop leaves the area invariant as a consequence of its symplectic nature. 


\textit{Beyond SQL sensing and detection---} To achieve beyond SQL sensing using the geometrical phase, (i) we need to ensure that the loop is closed $\oint dr = 0$, (ii) importantly the resulting phase should contain the force as a relative branch dependent detectable part and finally (iii) the squeezing should enhance this detectable part. To see very generally, it is useful to decompose the oscillator trajectory $\bm r(t) = \bm r_0(t) + s(t) \bm r_1(t)$, where $\bm r_0$ is the qubit independent force term ($\propto \eta$ in our example) and $\bm r_1$ is the qubit or branch dependent term ($\propto \gamma$ in our example). Then the detectable area dependent on the qubit branch $s$ is  
\begin{align}
    \Phi_\text{det} = \oint s \bm{r}_0^T J d\bm{r}_1 + \oint \bm r_0^T J \bm r_1 ds \;,
\end{align}
which contains the cross term between unconditional and conditional displacement. This cross term needs to be amplified to achieve beyond SQL sensing. In our protocol below, we apply squeezing to amplify $\bm r_0$ along $x$ and then apply $\pi$-pulses which effectively engineers a conditional displacement $\bm r_1$ along $p$ axis orthogonal to $\bm r_0$. We repeat the process to close the loop and obtain a detectable geometrical phase that contains the force and the enhancement due to squeezing, satisfying all of the conditions above.

}

\subsection{Geometric Protocol}
 We now describe our geometric sensing protocol that encodes the signal in geometric phase that depends on the area traversed by the oscillator in its phase space. We follow the scheme presented in the main text in Fig.~2(a), with the protocol divided into four parts. 
 We analyze each part of the cycle separately.
\begin{enumerate}
    \item {Time: $0-\tau/2$}. We have the total evolution as the free evolution of the oscillator conjugated by the squeezing operation given by $U_{1,s}(t,t_0) = \S^\dag(r e^{i2\omega t})U_s(t,t_0)\S(r e^{i2\omega t_0})$, where the squeezing phase comes from being in the rotating frame. We obtain
    \begin{align}
        U_{1,s} = \S^\dag(r) U_s(\tau/2, 0)\S(r) = e^{i \frac{\phi_0}{4}s } \D(2 e^r\alpha_s)
    \end{align}
where $\phi_0 = 8\pi \gamma \eta/\omega^2$.

\item Time: $\tau/2-3\tau/2$. We now have free evolution with $\pi$ pulses on the qubit, effectively reversing the sign of the spin. The total unitary for the second step is therefore given by 
\begin{align}
\begin{split}
    U_{2,s} &= U_s(3\tau/2,5\tau/4)U_{-s}(5\tau/4,3\tau/4) U_{s}(3\tau/4, 3\tau/2)    \\
    &=\bigg(e^{is\phi_0/4} \D(i\alpha_s) \D(-\alpha_s) \bigg) \bigg( e^{-is\phi_0/2}\D(-i\alpha_{-s})\D(-i\alpha_{-s})\bigg) \bigg(e^{is\phi_0/4} \D(\alpha_s) \D(i\alpha_s) \bigg) \\
&=\D(\alpha_s)\D({i}\beta_s)\D^\dag(\alpha_s) \;,
    \end{split}
\end{align}
where ${\beta_s = 2(\alpha_s-\alpha_{-s})=4\gamma s/\omega}$.

\item Time: $3\tau/2-2\tau$. Similar to step 1, we again conjugate the free evolution with squeezing and anti-squeezing, therefore, the unitary for general time is given by $U_{3,s}(t,t_0)= \S^\dag(r e^{i2\omega t})U_s(t,t_0)\S(r e^{i2\omega t_0})$, obtaining

\begin{align}
    U_{3,s} &=\S^\dag(r) U_s(2\tau, 3\tau/2)\S(r) = e^{i \frac{\phi_0}{4}s } \D^\dag(2 e^r\alpha_s) \;.
\end{align}

\item Time: $2\tau-3\tau$. Similar to step $2$, we again have $\pi$ pulses reversing the sign of spin. The total unitary for step 4 is given as 
\begin{align}
    \begin{split}
        U_{4,s} &= U_s(3\tau,11\tau/4)\ U_{-s}(9\tau/4,11\tau/4)\ U_s(9\tau/4,2\tau) \\
     &=   \bigg(e^{is\phi_0/4} \D(-i\alpha_s) \D(\alpha_s) \bigg) \bigg( e^{-is\phi_0/2}\D(i\alpha_{-s})\D(i\alpha_{-s})\bigg) \bigg(e^{is\phi_0/4} \D(-\alpha_s) \D(-i\alpha_s) \bigg) \\
&=\D^\dag(\alpha_s)\D^\dag({i}\beta_s)\D(\alpha_s) \;,
    \end{split}
\end{align}
\end{enumerate}

The total evolution is given by the unitary
\begin{align}
\begin{split}
        U_{T,s} &= U_{4,s} U_{3,s} U_{2,s} U_{1,s} \\
        &= e^{is\phi_0} \bigg(\D^\dag(\alpha_s)\D^\dag({i}\beta_s)\D(\alpha_s)\bigg) \D^\dag(2e^r \alpha_s) \bigg(\D(\alpha_s)\D({i}\beta_s)\D^\dag(\alpha_s)\bigg) \D(2e^r \alpha_s) \\
        &= e^{i\frac{\phi_T}{2}} \hat{I} 
        \end{split}
\end{align}
where the phase of our geometric protocol is $\phi_T = \phi_0 s + 4 A_s$, where the area $A_s = 2\alpha_s (e^r-1) \beta_s = 8(e^r-1)s \gamma\eta/\omega^2$ is geometrical phase equal to the area of the loop traversed in the oscillator phase space. Mathematically, the last equation can be simplified by using commutation of displacements $\D(\alpha)\D(\beta) = e^{i\Im(\alpha \beta^*)}\D(\alpha + \beta)$. Thus, the phase collected after $n$ loops is $n\phi_T = 8n (\pi + 4(e^r-1))\gamma \eta/\omega^2$. Using Eq.~\ref{eqn:qfi} to calculate the QFI, we obtain
\begin{align}
    \mathcal F_\eta = 64 n^2\frac{\gamma^2}{\omega^4}(\pi + 4(e^r-1))^2 \;,
    \label{eqn:QFI_geo}
\end{align}
where each loop takes $3\tau$ time period.  

\subsection{Squeezing with frequency change}
One of the experimentally feasible ways of obtaining squeezing is to suddenly change the oscillator frequency from $\omega$ to $\omega^\prime$. The Hamiltonian changes from $H =\omega \a^\dag \a$ to $H^\prime = \omega^\prime \a^{\prime\dag} \a^\prime$, where the annihilation and creation operators undergo a Bogoliubov transformation 
$a^\prime= \S^\dag(r) a \S(r)$, where the squeezing strength is $r = \ln(\omega/\omega^\prime)/2$. 
We now analyze our geometric protocol when the squeezing operation is applied using this sudden frequency change as shown in main text in Fig.~3(a). The Hamiltonian in the lab frame is now given as $H_L/\hbar = -\frac{\omega_q}{2} \sz + \omega_t \a^\dag_t \a_t + (\adag_t + \a_t) (\eta + \gamma s)$, where  $\a_t = \a, \a^\prime$ and $\omega_t = \omega, \omega^\prime$ for appropriate times $t$ consistent with Fig.~3(a). 
Since, the Hamiltonian and operators change piecewise, we move to the rotating frame $U_0 = e^{-i\frac{\omega_q t}{2}\sz} e^{i\theta_t \a^\dag_t \a_t}$, where $\theta_t = \int_0^t dt^\prime \omega_{t^\prime}$ is the total oscillator phase collected at time $t$. Using the transformation, $H =U_0 H_L U_0^\dag - i U_0 \dot U_0^\dag $, the Hamiltonian in the rotating frame is
\begin{align}
    H_t = (\a_t e^{-i\theta_t} + \adag_t e^{i\theta_t}) (\eta+\gamma s) \;.
\end{align}
When the frequency is only changed in a piecewise manner, the qualitative dynamics is the same as before, with the phase and displacement change appropriately with the frequency. For instance, free evolution of the oscillator with $\omega^\prime$ frequency is given by $U_s^\prime(t,0)=e^{is\phi^\prime(t)/2}\D(\alpha_s^\prime(t))$, where $\phi^\prime(t) = 4\eta \gamma (\omega^\prime t - \sin\omega^\prime t)/\omega^{\prime{2}}$ and $\alpha_s^\prime(t) = \alpha_s^\prime (1-e^{i\omega^\prime t})$ and $\alpha_s^\prime = (\eta+\gamma s)/\omega^\prime$. Since $\omega^\prime = \omega e^{-2r}$, the phase collected at the end of one cycle is four times that with frequency $\omega$: $\phi_0^\prime = \phi_0 e^{4r}$ and the displacement is amplified $\alpha_s^\prime = \alpha_s e^{2r}$. The second and fourth steps remain unaltered, while the first and third steps become
\begin{enumerate}
    \item Time: $0-\tau/2$. We have the free evolution of the oscillator with frequency $\omega^\prime$, conjugated by squeezing operation coming from the sudden change $ U_{1,s} =  \S^\dag(r e^{i2\omega t})U_s^\prime \S(r e^{i2\omega t_0})$, obtaining
    \begin{align}
       U_{1,s} = e^{i\frac{\phi_0 }{4}e^{4r}s}\D(2\alpha_s e^{3r}) \;.
    \end{align}
    \item Time: $3\tau/2-2\tau$. Similarly we obtain
    \begin{align}
        U_{3,s} = e^{i\frac{\phi_0 }{4}e^{4r}s}\D^\dag(2\alpha_s e^{3r}) \;.
    \end{align}
\end{enumerate}

The total evolution is similarly given by
\begin{align}
    U_{T,s} = e^{i\frac{\phi_T}{2}s} \hat{I} \;,
\end{align}
where the total phase is $\phi_T = e^{4r}\phi_0 + 4A_s$, where the area is $A_{{s}} = 2\alpha_s (e^{3r}-1)\beta_s = 8 (e^{3r}-1){s}\eta \gamma/\omega^2 + {\text{global}}$. Substituting this into the QFI expression, 
\begin{align}
\mathcal F_\eta = 64 n^2\frac{\gamma^2}{\omega^4}(e^{4r}\pi + 4(e^{3r}-1))^2 \;,\end{align}
and the time period per evolution is given by $T_g = \frac{2\pi}{\omega}\times 2 + \frac{2\pi}{\omega^\prime} = \frac{2\pi}{\omega} (2 + e^{2r})$.

\section{Dispersively Coupled Qubit-Oscillator Systems}
In typical cavity QED systems, the qubit is transversally coupled to the cavity. In the dispersive limit, the qubit-resonator coupling is {much} smaller than {their frequency difference}, with the effective Hamiltonian given by 
\begin{align}
    H= \frac{\chi}{2} \sz \adag\a + i \epsilon(t) \adag -i \epsilon^*(t) \a \;,
\end{align}
where $\chi$ is the dispersive shift and $\epsilon(t)$  is the drive on the resonator.
When there is no drive/pulse on the resonator, the free evolution is given by $\R_\theta = e^{-i s \frac{\theta}{2} \adag \a}$, where $\theta = \chi t$. We can drive the resonator by applying a pulse $\epsilon(t)$, displacing it by $\D(\alpha)$, where $\alpha = \int dt^\prime \epsilon(t^\prime)$. We now consider the following operations
\begin{enumerate}
    \item First, we implement a drive conjugated by free evolution
    \begin{align}
        U_{1} = \R_\theta^\dag \D(\alpha) \R_\theta = \D(\alpha e^{is\frac{\theta}{2}}) \;,
    \end{align}
    where $\R_\theta^\dag$ is implemented by conjugating the free evolution $\R_\theta$ with two $\pi$ pulses, reversing the sign of qubit.
    \item Next, we have a drive implementing the real displacement, conjugated by squeezing
    \begin{align}
        U_2 = \S^\dag(r) \D(\alpha) \S(r) = \D(\alpha e^r) \;.
    \end{align} 
    \item We {reverse} the first unitary by applying a drive with $\pi$ phase 
    \begin{align}
        U_3 = \R_\theta^\dag \D(-\alpha) \R_\theta = \Ddag(\alpha e^{is\frac{\theta}{2}}) \;.
    \end{align}
    \item Similarly, we {reverse} the second step, completing the loop 
    \begin{align}
        U_4 = \S^\dag(r) \D(-\alpha) \S(r) = \Ddag(\alpha e^r) \;.
    \end{align}
\end{enumerate}
    
Finally, we obtain the total evolution 
\begin{align}
    U = U_4 U_3 U_2 U_1 = e^{i \phi_s /2} \;,
\end{align}
where the geometric phase $\phi_s = 4 \alpha^2 e^{r} \sin(s \chi t/2)$. Since $s = \pm 1$, we can just simplify the expression to the form we have been using in the paper 
\begin{align}
    U = e^{i s \phi/2}, 
\end{align}
where $\phi = 4 \alpha^2 e^{r} \sin(\chi t/2)$. We can now use our standard formula to obtain the QFI for both the displacement which can be used to obtain the pulse amplitude and the dispersive coupling after $n$ loops 
\begin{align}
    \mathcal F_\alpha = (\partial_\alpha \phi)^2 = 64 n^2 \alpha^2 e^{2r} \sin^2(\chi t/2) \;,
\end{align}
\begin{align}
    \mathcal F_\chi = (\partial_\chi \phi)^2 = 4 n^2 t^2 \alpha^4 e^{2r} \cos^2(\chi t/2) \;.
\end{align}

{
\section{Sensitivity to errors}
In reality, noise from the environment or control pulse errors can lead to both residual
entanglement between qubit and oscillator and an error in the geometric phase $\phi(t)$. Here we
give bounds on each to achieve a desired sensitivity $\epsilon$. A $\sx$ measurement on the qubit
will generally lead to the signal $S \equiv \langle \sx \rangle=e^{-d} \cos\Phi$, where $d$ is the
residual entanglement and $\Phi = \phi_T + \delta \phi$, with the ideal phase
$\phi_T = \eta \sqrt{\mathcal F_\eta} = 8 \eta \gamma(\pi + 4 (e^r - 1))/\omega^2$ from
Eq.~\ref{eqn:QFI_geo} and $\delta \phi$ the pulse error. Here $\mathcal F_\eta$ denotes the ideal
QFI, the damping being retained explicitly. The force is then estimated as
$\eta = (\arccos (S e^d) - \delta \phi)/\sqrt{\mathcal F_\eta}$, where $d$ and $\delta\phi$ are
known to within calibration uncertainties $\Delta d$, $\Delta\delta\phi$. Using the error
propagation formula, we have
\begin{align}
    (\Delta \eta)^2 = \frac{1}{\mathcal F_\eta \sin^2 \Phi}
    \Big(e^{2d} \Delta S^2 + \cos^2 \Phi\, \Delta d^2 + \sin^2 \Phi\, \Delta \delta \phi^2\Big) \;.
    \label{eq:prop}
\end{align}
For a binary readout after $\nu$ measurements, $\Delta S^2 = (1-S^2)/\nu =
(1-e^{-2d}\cos^2\Phi)/\nu$, whereupon the entire first term reduces to
$(e^{2d}-\cos^2\Phi)/(\nu \mathcal F_\eta \sin^2\Phi) = 1/(\nu \mathcal F^c_\eta)$, i.e.\ error
propagation independently reproduces the classical Fisher information of Eq.~\ref{eqn:cfi} as its
statistical contribution, which serves as a consistency check and identifies the remaining two
terms as the systematic ones. Furthermore since every term is divided by $F_\eta$, squeezing suppresses both, the statistical and the systematic errors $\Delta d, \Delta \delta \phi$. Demanding $\Delta \eta < \epsilon$, implies $\nu \geq 1/(F_\eta^c \epsilon^2)$, $\Delta d \leq \epsilon \sqrt{F_\eta} |\tan \Phi|$ and $\Delta \delta \phi \leq \epsilon\sqrt{F_\eta}$.

\section{Choice of noise parameter $\lambda$ for numerical simulation}


The noise parameter $\lambda$ is normalized with respect to the oscillator frequency $\omega$ in our numerical simulations (Fig.~5) and spans several orders of magnitude, reflecting differences between mechanical and electromagnetic platforms. Modern transmon qubits exhibit coherence times $T_1 \in {100~\mu\text{s} - 1~\text{ms}}$~\cite{bland2025transmon_1ms}, corresponding to qubit noise levels $\lambda_{\rm qubit} = 1/(\omega T_1) \in {0.1 - 0.01}$ for mechanical oscillators ($\omega \sim 10^{5}$) and $\lambda_{\rm qubit} \in {10^{-7} - 10^{-6}}$ for electromagnetic oscillators ($\omega \sim 10^{10}$). Similarly, mechanical (electromagnetic) resonators routinely achieve quality factors exceeding $10^{4}$ ($10^{8}$)~\cite{johnsson2016macroscopic, kim2025ultracoherent, oriani2025Nb}, corresponding to oscillator noise levels $\lambda_{\rm osc} = Q^{-1} \lesssim 10^{-4}$ ($10^{-8}$), where $Q=\omega/\kappa$ and $\kappa$ is the oscillator bandwidth.
}

\bibliography{Sensing}